\documentclass[9pt,letterpaper]{article}
\usepackage{opex3}
\usepackage{cite}

\begin{document}
\title{Imprints of the molecular-orbital geometry on the high-harmonic ellipticity }
\author{Meiyan Qin,$^{1}$ Xiaosong Zhu,$^{1}$ Kunlong Liu,$^{1}$ Qingbin Zhang,$^{1,2,3}$ and Peixiang Lu$^{1,2}$$^*$}
\address{$^1$Wuhan National Laboratory for Optoelectronics and School of Physics, Huazhong University of Science and Technology, Wuhan 430074, China \\
$^2$MOE Key Laboratory of Fundamental Quantities
Measurement,Wuhan430074,China\\
$^3$zhangqingbin@mail.hust.edu.cn}
\email{$^*$lupeixiang@mail.hust.edu.cn}

\begin{abstract}
The influence of the orbital symmetry on the ellipticity of the
high-order harmonics is investigated. It is found that the
ellipticity maps have distinct shapes for the molecular orbitals
with different symmetry. Our analysis shows that the feature of
the harmonic ellipticity map is essentially determined by the
nodal structure of the nonsymmetric orbital. The results indicate
that the molecular-orbital geometry is imprinted on the
ellipticity of the high-order harmonics, which invites the use of
ellipticity measurements as a probe of the orbital structure for
polar molecules.
\end{abstract}


\section{Introduction}
Progress in strong-field physics has made it possible to probe the
molecular structure and dynamics with attosecond and
$\mathrm{\AA}$ngstr\"{o}m resolutions \cite{lein,lein1,liao,hong}.
High-order harmonic generation (HHG), which occurs in the
interaction of strong field with atoms or molecules, is one of the
most important tool for probing the molecular structure and
dynamics. In particular, the application of HHG in the molecular
orbital tomography (MOT) and attosecond probing of electronic
dynamics in molecules has attracted a great deal of attention in
the past decade
\cite{itatani,gibson,zwan,haessler,ramakrishna,mairesse,smirnova}.
Combined with molecular alignment techniques, the reconstruction
of the highest occupied molecular orbital (HOMO) of N$_{2}$ was
demonstrated in 2004 \cite{itatani}. The studies of MOT have shown
that the orbital symmetry plays an important role in the
reconstruction of the molecular wavefunction
\cite{itatani,gibson,zwan,haessler}. Therefore, the prior
knowledge of the orbital symmetry is essential to the molecular
orbital tomography based on the HHG.

In previous works, several methods were proposed to probe the
orbital symmetry \cite{levesque,zhou,hijano,levesque2,niikura}.
For example, the information of the orbital symmetry can be
decoded from the polarization angle of the high-order harmonics
with respect to the driving linearly polarized laser field
\cite{levesque,zhou,hijano}. Another method to probe the orbital
symmetry is to measure the alignment-dependent harmonic spectrum
\cite{levesque2,niikura}. In these works, the investigated
molecular states are symmetric with defined parity (gerade or
ungerade). Whereas for nonsymmetric orbitals that are neither
gerade nor ungerade, obtaining the information about the orbital
geometry has not been specifically discussed. In our recent work,
it has been shown that for nonsymmetric molecular states, the
ellipticity of the high harmonics are quite large in a wide
spectral range, and this polarization property is mainly
attributed to the nonsymmetric structure of the orbital
\cite{qin}. Thus, it would be an effective method to obtain the
information of the orbital geometry from the ellipticity of the
high-order harmonics for nonsymmetric molecular states.

In this paper, we investigate the influence of the orbital
symmetry on the ellipticity of the high-order harmonics generated
from nonsymmetric molecular states. It is found that the
ellipticity maps of the high-order harmonics have unique shapes
for orbitals with different symmetry. In order to gain a deeper
insight into the features of the harmonic ellipticity, structural
analysis of these molecular orbitals is performed. In addition,
the individual contributions of s and p atomic states of the
molecular orbital to the ellipticity of the high-order harmonics
are also discussed.

\section{Theoretical model}
The molecular orbitals are simulated by using the linear
combination of atomic orbitals (LCAO) approximation, along with
the Born Oppenheimer approximation \cite{faria,augstein}. Under
these assumptions, the molecular orbital wavefunction is given by
\begin{eqnarray}
\psi_{0}(\mathbf{r})=\sum_{n}c_{1}^{n}\varphi_{1}^{n}(\mathbf{r}-\mathbf{R}_{1})
+c_{2}^{n}\varphi_{2}^{n}(\mathbf{r}-\mathbf{R}_{2})
\end{eqnarray}
In this equation, the sum over n denotes the sum over the atomic
orbitals. The number is used to distinguish the different
molecular centers. The internuclear separation is given by
$R=\mid\mathbf{R}_{1}-\mathbf{R}_{2}\mid$. All these parameters
are extracted from the Gaussian 03 ab initio code \cite{gauss}.
Gaussian-type orbitals with the 6-311G basis set are employed. The
z-axis in the lab frame is defined as the polarization axis of the
linearly polarized laser. The molecular axis is rotated in the
$y$-$z$ plane. The orientation angle between the internuclear axis
and the z axis is denoted by $\theta$. The driving laser field
propagates along the x-axis.

A widely used quantum theory to model the process of the high
harmonic generation is the strong field approximation (SFA)
\cite{corkum,lewen,lan1}. With its analytical and fully
quantum-mechanical formulations for high harmonics, one can make a
clear interpretation of the results in terms of classical physics.
It has been shown that the expression for the HHG field can be
factorized into two terms: the recombination dipole matrix element
and a complex continuum electron wave packet (EWP) amplitude, when
extremely short laser pulse is applied
\cite{zwan,levesque2,haessler}. Accordingly, the HHG field of the
components parallel and perpendicular to the polarization axis of
the driving field are given by (in atomic units)

\begin{equation}
A_{z}(\omega,\theta)=2\pi\omega W(E_{k},\theta)d_{z}(k;\theta)
\end{equation}
\begin{equation}
A_{y}(\omega,\theta)=2\pi\omega W(E_{k},\theta)d_{y}(k;\theta)
\end{equation}
In these equations, $\omega$ is the frequency of the high harmonic
emission, and $E_{k}=k^{2}/2$ is the kinetic energy of the
returning electron. They are related through $\omega=E_{k}+I_{p}$,
with $I_{p}$ being the ionization energy of the state that the
electron ionized from. $W(E_{k},\theta)$ stands for the returning
electron wave packet (EWP) amplitude. The complex wave packet in
eq. (2) and eq. (3) are identical, since the returning electron is
the same for the two components of the harmonic field. $d_y, d_z$
are the $\mathbf{y}$, $\mathbf{z}$ components of the recombination
dipole moment respectively. The recombination dipole moment is
expressed as,
\begin{equation}
\vec{d}(k,\theta)=\langle\psi_{0}(x,y,z;\theta)|\vec{r}|e^{ikz}\rangle
\end{equation}
In this equation, $\psi_0(x,y,z;\theta)$ is the investigated
molecular orbital wavefunction given by equation (1). Using Eqs.
(2)-(4), one can obtain the amplitude ratio and the phase
difference of the two components of the high harmonics by $r=|A_y
/ A_z|=|d_y/d_z|$ and $\delta = \arg[A_y]-\arg[A_z]=
\arg[d_y]-\arg[d_z]$. The ellipticity of the high-order harmonics
$\epsilon$ as a function of $r$ and $\delta$ is given by
\cite{son}:
\begin{equation}
\epsilon =
\sqrt{\frac{1+r^2-\sqrt{1+2r^{2}\cos2\delta+r^4}}{1+r^2+\sqrt{1+2r^2\cos2\delta+r^4}}}
\end{equation}
The range of the ellipticity is $0\leq\epsilon\leq1$. The linear,
elliptical, and circular polarization correspond to $\epsilon=0$ ,
$0<\epsilon<1$ , and $\epsilon=1$ respectively. When $r=1$ and
$\delta = \pi/2$, circularly polarized high-order harmonics with
$\epsilon=1$ will be obtained.

\section{Result and discussion}

Four typical types of nonsymmetric orbitals are considered:
$\pi$-type bonding and anti-bonding orbitals, and $\sigma$-type
bonding and anti-bonding orbitals. These orbitals correspond to
four basic orbital symmetries which are denoted by $\pi$,
$\pi^{\ast}$, $\sigma$ and $\sigma^{\ast}$, respectively. Our
study mainly focuses on the influence of the orbital symmetry on
the ellipticity of the high-order harmonics. Hence the model
molecules with similar internuclear distances (NO and CO) are
used. The $\pi^{\ast}$ highest occupied molecular orbital (HOMO)
of NO and the $\pi$ HOMO-1 of CO are considered in our
investigation. For comparison, the parameters $c_{1}^{n},
c_{2}^{n}, \mathbf{R}_{1}, \mathbf{R}_{2}$ of the HOMO-1 of CO and
the HOMO of NO are also used to construct $\sigma^{\ast}$ and
$\sigma$ orbital wavefunctions, respectively. The parameters
$c_{1}^{n}, c_{2}^{n}$ of the HOMO of NO and the HOMO-1 of CO are
listed in Table 1. With Eqs. (2)--(5), the ellipticity of the
high-order harmonics are obtained for these molecular states. The
results are presented in the upper panels of Fig. 1. The
horizontal axis stands for the orientation angle, and the vertical
axis stands for the kinetic energy of the returning electron. The
range of the kinetic energy is from 5eV to 40eV. The sectional
views of the molecular orbitals are also presented in Fig. 1 (the
lower panels). The orientation angle $\theta$ is $0^{\circ}$ for
$\pi$, $\sigma$ and $\sigma^{\ast}$ orbitals, and is $50^{\circ}$
for $\pi^{\ast}$ orbital. The nodal structures of molecular
orbitals are depicted by the white dashed lines.

\begin{center}
\small
\begin{tabular} {|c|c|c|c|c|c|c|c|c|}
 \multicolumn {9}{c}
 {\bfseries TABLE 1: The parameters $c_{1}^{n}, c_{2}^{n}$ used in our simulation } \\
\hline
 \hline
  & $c_{1/2}^{1s}$ & $c_{1/2}^{2s}$ & $c_{1/2}^{3s}$ & $c_{1/2}^{2p_{y}}$ & $c_{1/2}^{2p_{z}}$ & $c_{1/2}^{3p_{y}}$ & $c_{1/2}^{3p_{z}}$ & $c_{1/2}^{4p_{y}}$ \\
 \hline
  HOMO & -0.151 & 0.134 & 0.776 & $\setminus$ & -0.384 & $\setminus$ & -0.174 & $\setminus$  \\
   of CO  & 0.013 &-0.016 & -0.094 & $\setminus$ & 0.221 & $\setminus$ &0.241 & $\setminus$ \\
  \hline
  HOMO-1& $\setminus$ & $\setminus$ & $\setminus$ & 0.252 & $\setminus$ & 0.199& $\setminus$ & $\setminus$   \\
     of CO & $\setminus$ &  $\setminus$ & $\setminus$  & 0.446 & $\setminus$ & 0.486 & $\setminus$ & $\setminus$   \\
  \hline
  HOMO & $\setminus$ & $\setminus$ & $\setminus$  & 0.255 & $\setminus$ & 0.413 & $\setminus$ & 0.402   \\
   of NO  & $\setminus$ & $\setminus$  & $\setminus$  & -0.234 & $\setminus$ &-0.358 & $\setminus$ &-0.402 \\
 \hline
 \end{tabular}
 \end{center}
\begin{figure}[htd]
\centerline{
\includegraphics[width=9cm]{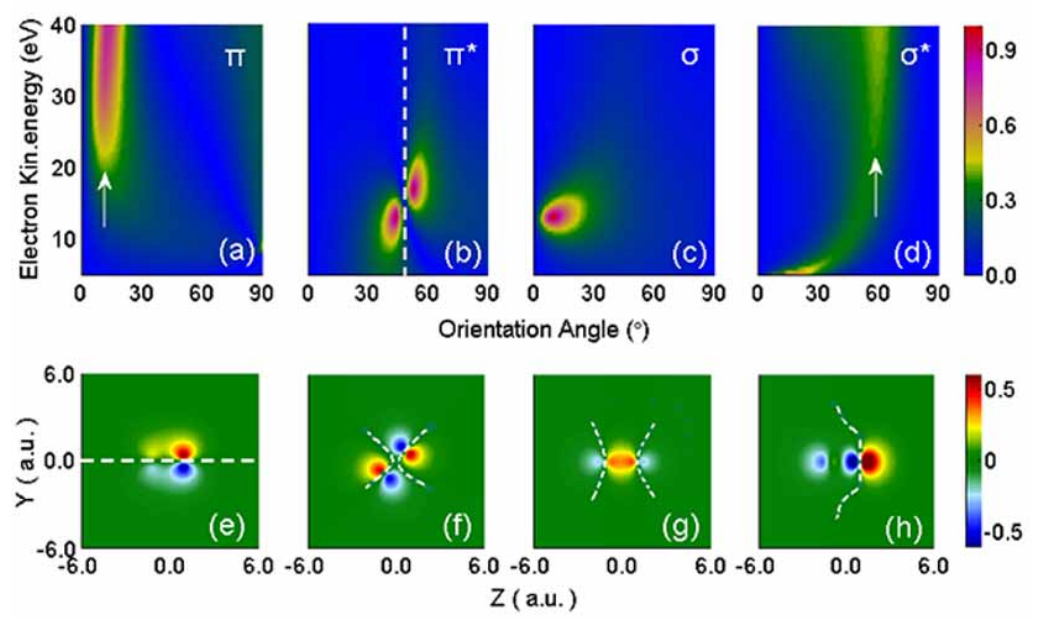}}
\caption{The ellipticity maps for the $\pi$ (panel a),
$\pi^{\ast}$ (panel b), $\sigma$ (panel c) and $\sigma^{\ast}$
(panel d) orbitals . The horizontal axis corresponds to the
orientation angle, the vertical axis corresponds to the kinetic
energy of the returning electron. In panels (a) and (d), the
arrows indicate the maxima of the ellipticity for high kinetic
energy (above 20eV). The sectional views of the corresponding
molecular orbitals are presented in the second row. The
orientation angle is 0$^{\circ}$ for the $\pi$, $\sigma$ and
$\sigma^{\ast}$ orbitals, and is $50^{\circ}$ for the $\pi^{\ast}$
orbital. The nodal structures of molecular orbitals are depicted
by the white dashed lines.}
\end{figure}

\begin{figure}[b]
\centerline{
\includegraphics[width=11cm]{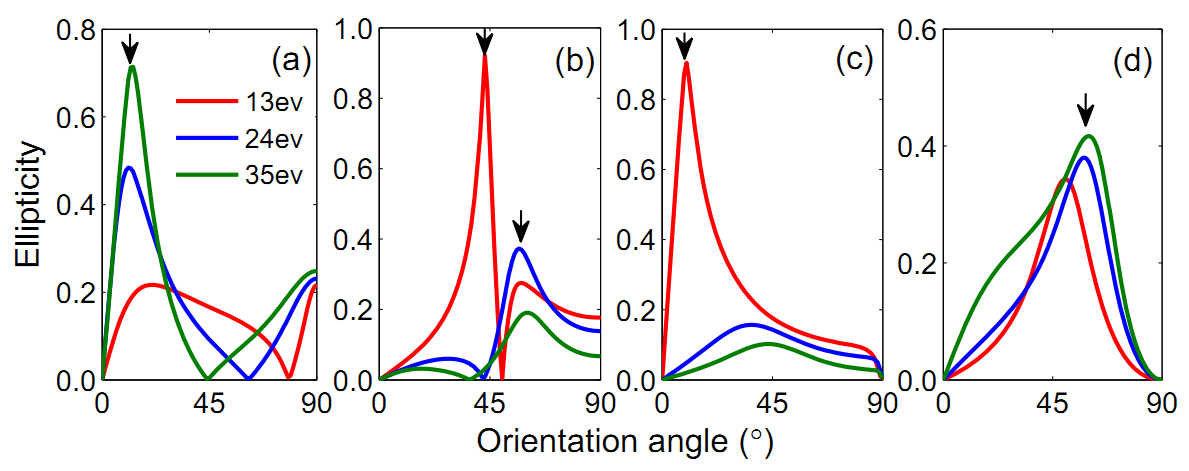}}
\caption{The ellipticity as a function of the orientation angle
for three given kinetic energies: 13eV (the red curve), 24eV (the
blue curve), and 35eV (the green curve). Panels (a), (b), (c) and
(d) correspond to the $\pi$, $\pi^{\ast}$, $\sigma$ and
$\sigma^{\ast}$ orbitals, respectively. The black arrow indicates
the maximum of the harmonic ellipticity.}
\end{figure}
As shown in Fig. 1, the ellipticity maps have distinct shapes for
$\pi$, $\pi^{\ast}$, $\sigma$ and $\sigma^{\ast}$ orbitals. In
Fig. 1(a) for the $\pi$ orbital, large ellipticity of the high
harmonics are observed at small angles. Moreover, the maxima of
the ellipticity appear almost at the same orientation angle of
$12^{\circ}$ for the kinetic energies above 20eV, as indicated by
the white arrow. In panel (b) for $\pi^{\ast}$ orbital, the
ellipticity map of the high-order harmonics is divided into two
parts with similar shapes by the white dashed line at
$50^{\circ}$. In each part, one peak of the ellipticity of the
high harmonics is observed at the orientation angle close to
$50^{\circ}$. In panel (c) for $\sigma$ orbital, one peak of the
ellipticity of the high-order harmonics appears at the orientation
angle close to $0^{\circ}$. Interestingly, the shape of the
ellipticity map in Fig. 1(c) is quite similar to the shape of the
right part in Fig. 1(b). In the case of $\sigma^{\ast}$ orbital,
the maxima of the ellipticity of the high harmonics are observed
at large orientation angles around $60^{\circ}$ for all the
kinetic energy higher than 20eV, as indicated by the arrow in Fig.
1(d).
\begin{figure}[htd]
\centerline{
\includegraphics[width=11cm]{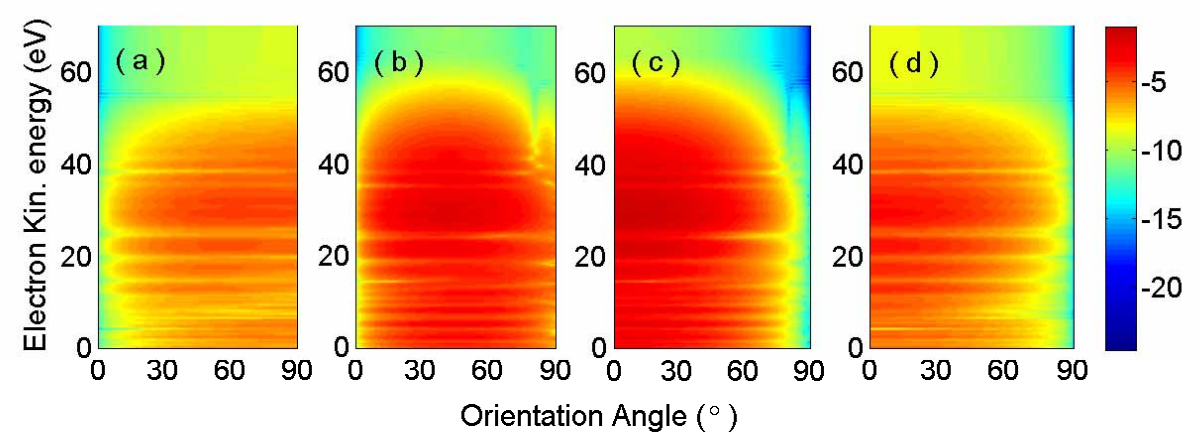}}
\caption{The harmonic spectra for $\pi$ ( a ), $\pi^{\ast}$ ( b ),
$\sigma$ ( c ) and $\sigma^{\ast}$ ( d ) orbitals. A three-cycle
sin$^2$ pulse with a carrier-envelope phase of 1.25 $\pi$ is used.
The peak intensity and wavelength of the laser field are
$2\times10^{14}$ $\mathrm{W/cm^2}$ and 800 nm, respectively. The
horizontal axis corresponds to the orientation angle, and the
vertical axis corresponds to the kinetic energy of the returning
electron.}
\end{figure}

\begin{figure}[b]
\centerline{
\includegraphics[width=11cm]{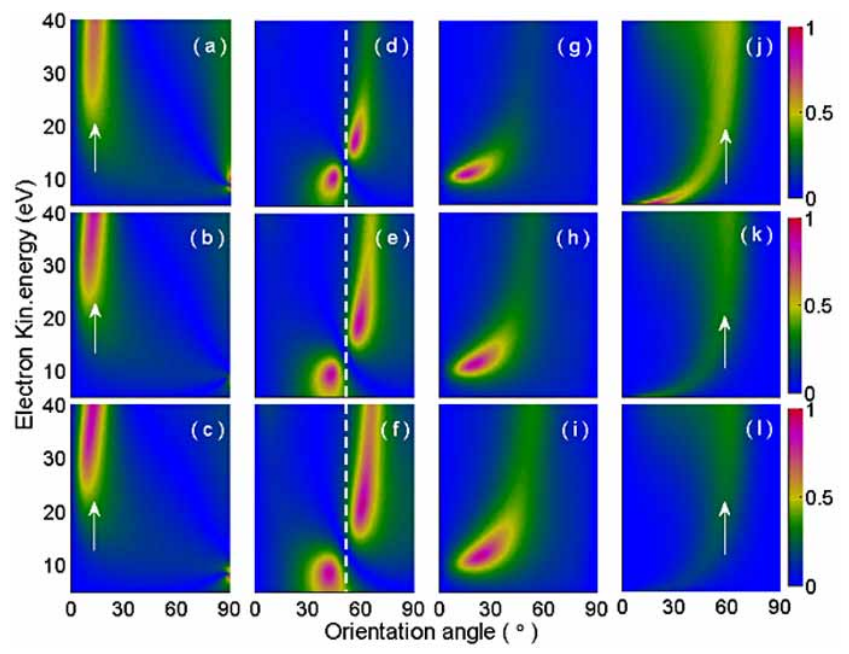}}
\caption{The ellipticity maps of the high-order harmonics for the
$\pi$ (the first column), $\pi^{\ast}$ (the second column),
$\sigma$ (the third column) and $\sigma^{\ast}$ (the fourth
column) orbitals with various weight of the contribution from
nucleus 1 to the molecular orbital. From the first row to the
third row, the weight of the contribution from nucleus 1
increases.}
\end{figure}

For clarity, the ellipticity as a function of the orientation
angle for three given kinetic energies are presented in Fig. 2 for
$\pi$, $\pi^{\ast}$, $\sigma$ and $\sigma^{\ast}$ orbitals. The
black arrows indicate the maxima of the ellipticity of the high
harmonics. As shown in Fig. 2(a) for $\pi$ orbital, large
ellipticity of the high harmonics are obtained mainly at small
orientation angles. The maxima of the ellipticity for high kinetic
energy 24eV (the blue curves) and 35eV (the green curves) are
almost fixed at a small orientation angle of $12^{\circ}$. While
in panel (d) for $\sigma^{\ast}$ orbital, these maxima appear
around a large orientation angle of $60^{\circ}$. In panel (b) for
$\pi^{\ast}$ orbital, large ellipticity of the high harmonics are
observed at intermediate angles. The ellipticity of the high-order
harmonics at a given kinetic energy has two peaks around
$50^{\circ}$. In panel (c) for $\sigma$, large ellipticity of the
high harmonics are observed at small orientation angles close to
$0^{\circ}$. The harmonic ellipticity at a given kinetic energy
has one peak.

In Fig. 3, the harmonic spectra versus both the kinetic energy of
the returning electron and the orientation angle are presented for
the four molecular states. A three-cycle $\mathrm{sin}^{2}$ pulse
with a carrier-envelope phase of 1.25$\pi$ is used, where the
factorization of the complex spectrum can be achieved \cite{zwan}.
The peak intensity and wavelength of the laser field are
$2\times10^{14}$ $\mathrm{W/cm^2}$ and 800 nm, respectively. The
high-order harmonics corresponding to the kinetic energy from 5eV
to 40eV are well located in the plateau of the spectra for these
molecular states. From Fig. 1 and Fig. 3, one can see that the
ellipticity measurements considered in Fig. 1 correspond to
measurable harmonic signals.

To gain a deeper insight into the features of the harmonic
ellipticity observed in Figs. 1(a)--1(d), the structural analysis
of the molecular state is performed. On the one hand, for the
$\pi$ and $\sigma^{\ast}$ orbitals shown in Fig. 1(e) and Fig.
1(g), there is only one obvious nodal surface. In Fig. 1(e) for
$\pi$ orbital, the nodal surface is parallel to the z axis of the
lab frame. Correspondingly, the maxima of the ellipticity for
$\pi$ orbital appear at small angles around $12^{\circ}$ for all
the kinetic energy above 20eV. In Fig. 1(h) for $\sigma^{\ast}$
orbital, the nodal surface is perpendicular to the z axis of the
lab frame disregarding the distortion of the nodal surface.
Corresponding to this direction of the nodal surface, the maxima
of the ellipticity of the high harmonics appear at large angles
around $60^{\circ}$ for all the kinetic energy higher than 20eV.
Therefore, for the molecular state with only one obvious nodal
plane, the maxima of the ellipticity of the high-order harmonics
in high-energy range appear at almost the same orientation angle.
This angle is determined by the direction of the nodal surface
with respect to the z axis of the lab frame.

On the other hand, for the $\pi^{\ast}$ and $\sigma$ orbitals
shown in Figs. 1(f)-1(g), there are two obvious nodal surfaces. In
addition, the shape and direction of the nodal surfaces for
$\pi^{\ast}$ at the orientation angle of $50^{\circ}$ are nearly
the same as those of the nodal surfaces for $\sigma$ at
$\theta=0^{\circ}$. As depicted by the white dashed lines, both
the nodal surfaces show a ``$> <$'' structure. Correspondingly,
the peaks with similar shapes are observed in the right part
($\theta>50^{\circ}$) of the ellipticity map for $\pi^{\ast}$ and
in the ellipticity map for $\sigma$. The location of the peak is
close to $50^{\circ}$ in the former case and close to $0^{\circ}$
in the latter case. Additionally, in Fig. 1(f) for the
$\pi^{\ast}$ orbital, the orbital at $\theta=50^{\circ}$ is
approximately symmetric with respect to the z axis. The rotation
of the nodal surfaces with $\theta$ varying from $50^{\circ}$ to
$0^{\circ}$ is similar to that with $\theta$ from $50^{\circ}$ to
$90^{\circ}$. Thus the left part of the ellipticity map with
$\theta$ varying from $50^{\circ}$ to $0^{\circ}$ has a similar
shape to the right part with $\theta$ varying from $50^{\circ}$ to
$90^{\circ}$ for $\pi^{\ast}$ orbital. Both in the two parts, the
peaks of the ellipticity of the high-order harmonics appear at the
orientation angles close to $50^{\circ}$. It is shown that the
nodal structures are imprinted on the ellipticity of the
high-order harmonics and consequently the molecular orbitals with
different symmetry correspond to distinct ellipticity maps.

To further confirm our conclusion, the ellipticity of the
high-order harmonics for the molecular orbitals with various
orbital distributions are investigated. These orbitals are
constructed by the same atomic states as those in Fig. 1 with
different weights of the contribution from the two nuclei. The
nodal structures are the same as those in Fig. 1. The ellipticity
maps of these molecular orbitals are presented in Fig. 4. From the
first row to the third row, the weight of the contribution from
nucleus 1 increases. Similar to those in Fig. 1, the results for
the $\pi$ and $\sigma^{\ast}$ orbitals with only one nodal surface
are presented in the first and last columns. As shown in Figs.
4(a)--4(c) for the $\pi$ orbitals, large ellipticity of the
high-order harmonics are observed at small angles, and the maxima
of the ellipticity appear almost at the same orientation angle.
The shapes of the ellipticity maps are similar to that in Fig.
1(a) and the location of the ellipticity maxima are also at around
$12^{\circ}$. In the case of $\sigma^{\ast}$ orbitals shown in the
last column, the features of the harmonic ellipticity are also the
same as that observed in Fig. 1(d), i.e. the maxima of the
ellipticity of the high-order harmonics in high kinetic energy
range locate around the large orientation angle of $60^{\circ}$.
From Fig. 4(j) to Fig. 4(l), one can see that the ellipticity of
the high-order harmonics decrease. This is because the orbital
distribution tends to be symmetric when the weight of the
contribution from nucleus 1 increases \cite{qin}. This phenomenon
is also observed in the case of the $\pi$ orbital when further
increasing the weight of the contribution from nucleus 1.

The ellipticity maps of the high-order harmonics for the
$\pi^{\ast}$ and $\sigma$ orbitals with the ``$> <$'' nodal
structure are presented in the second and third columns of Fig. 4.
In the case of the $\pi^{\ast}$ orbitals shown in Figs.
4(d)--4(f), all the ellipticity maps of the high-order harmonics
are divided into two parts by the white dashed line at
$50^{\circ}$. In each part, one peak of the ellipticity is
observed at the orientation angle close to $50^{\circ}$. These
features of the harmonic ellipticity are the same as those
observed in Fig. 1(b). From Fig. 4(d) to Fig. 4(f), the shapes of
the two parts of the ellipticity map become different. This is
because the degree of asymmetry for the wavefunctions at
$50^{\circ}$ with respect to the z axis increases when the
contribution of the atomic orbitals at nucleus 1 to the molecular
orbitals increases. In the third column of Fig. 4 for the $\sigma$
orbitals, one peak of the harmonic ellipticity is observed in each
panel, which is consistent with that presented in Fig. 1(c).
Therefore, in the four cases of the molecular orbitals with
different symmetry, the main features of the harmonic ellipticity
are not changed by varying the orbital distribution and are the
same as those observed in Fig. 1.

\begin{figure}[htb]
\centerline{
\includegraphics[width=10cm]{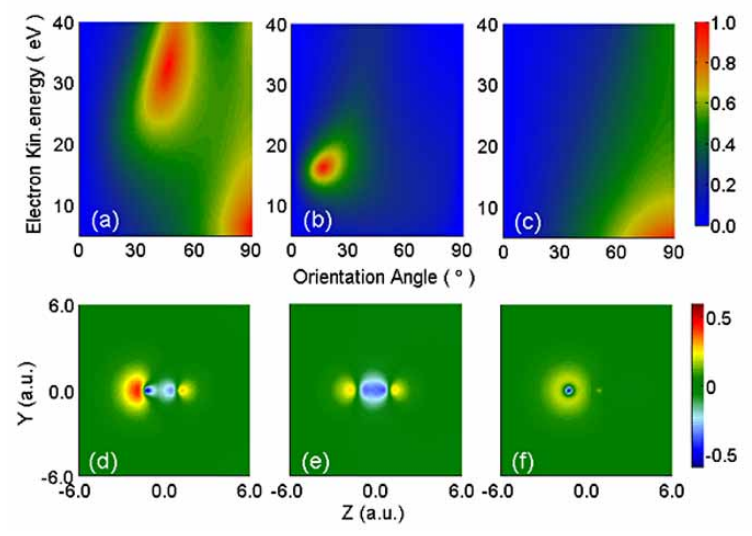}}
\caption{The ellipticity map for (a) the full orbital of the CO
HOMO, (b) the contribution from the p-type atomic states, (c) the
contribution from the s-type atomic states. The sectional views of
the corresponding wavefunctions are presented in lower panels. The
orientation angle is $0^{\circ}$ for all the wavefunctions.}
\end{figure}
The molecular orbitals considered in Fig. 1 are constructed from
the p-type atomic states. According to the simulation with the
Gaussian 03 ab initio code, the $\pi$ and $\pi^{\ast}$ orbitals
are constructed entirely from p-type atomic states, while the
$\sigma$ and $\sigma^{\ast}$ orbitals may be the mixing between s
and p atomic states. Both the symmetry and nodal structure of the
s atomic state are different from those of the p state. In the
following, we employ the $\sigma$-type HOMO of CO to investigate
the individual contributions of the s and p states to the
ellipticity of the high-order harmonics. The parameters
$c_{1}^{n}, c_{2}^{n}$ of the HOMO of CO are given in Table 1. As
in the references \cite{faria,augstein}, the parameters
$c_{1}^{n}, c_{2}^{n}$ of s (p) atomic states are set to zero when
investigating the individual contributions of the p (s) atomic
states. By comparing the ellipticity maps of the molecular
orbitals with and without the contribution of s atomic states, we
investigate the role played by the s-p mixing in the ellipticity
of the high-order harmonics.

The ellipticity map of the high-order harmonics generated from the
HOMO of CO is presented in Fig. 5(a). The individual contributions
of the p and s states are displayed in Fig. 5(b) and Fig. 5(c),
respectively. The sectional views of the corresponding
wavefunctions are also presented in the lower panel. As shown in
this panel, the HOMO of CO exhibits a strong mixing between s and
p states. The nodal structure of the full $\sigma$ orbital in Fig.
5(d) becomes a closed surface due to the mixing. The strong s p
mixing will leave an imprint on the ellipticity of the high-order
harmonics. In detail, the orbital wavefunction of the p states
presented in Fig. 5(e) shows the same nodal structure (``$> <$'')
as that in Fig. 1(g). Correspondingly, the shape of the
ellipticity map in Fig. 5(b) is similar to that in Fig. 1(c). In
the case of the s atomic states shown in Fig. 5(c), the
ellipticity map presents a new shape due to the new nodal
structure and large ellipticity of the high-order harmonics are
observed at the orientation angles around $90^{\circ}$. In Fig.
5(a) for the full $\sigma$ orbital, large ellipticity of the
high-order harmonics are obtained in two parts, due to the effect
of the strong s-p mixing. The first part locates at large
orientation angles around $90^{\circ}$ and exhibits similar shape
to that in Fig. 5(c) for the s states. The second part covers the
orientation angles from $30^{\circ}$ to $60^{\circ}$. Moreover,
similar to the ellipticity map in Fig. 5(b) for the p states,
there is also a peak in the second part. Accordingly, the harmonic
ellipticity in the first part is mainly contributed by the s
states, and the contribution of the p states dominates in the
second part. To confirm this, we also study the harmonic
ellipticity for the molecular orbitals with different weight of
the contribution of the s states. It is found that large
ellipticity observed at the orientation angles around $90^{\circ}$
decreases quickly when the weight of the s states is reduced. The
shape of the ellipticity map in Fig. 5(a) is changed into that
presented in Fig. 5(b) when the weight of the s states is reduced
to zero.

As shown in Fig. 1 and Fig. 5, the shapes of the ellipticity maps
for the $\sigma$ orbital (both with and without the s-p mixing)
are distinct to those of the ellipticity maps for the $\pi$,
$\pi^{\ast}$ and $\sigma^{\ast}$ orbitals. We also investigate the
effect of the s-p mixing on the ellipticity of the high-order
harmonics for the $\sigma^{\ast}$ orbital. In general, the nodal
structure of the $\sigma^{\ast}$ orbital is not significantly
changed by the s-p mixing due to the antibonding symmetry. Thus
for the $\sigma^{\ast}$ orbital (both with and without the s-p
mixing), the ellipticity map of the high-order harmonics has
similar shape to that presented in Fig. 1(d). The results indicate
that one can obtain the information of the orbital symmetry from
the ellipticity map of the high-order harmonics for the
nonsymmetric orbitals.

As discussed in Section 2, extremely short laser pulse is required
to achieve the factorization of the complex spectrum for
nonsymmetric orbitals, which benefits the application of the
harmonic ellipticity as a probe of the orbital structure. To
further investigate the role played by the pulse duration, we also
calculate the harmonic ellipticity map using the laser pulses with
10fs and 20fs duration, where the complex spectrum can not be
factorized in the recombination dipole moment and a complex
continuum EWP. It is found that the main features of the
ellipticity maps for $\pi$, $\pi^{\ast}$, $\sigma$ and
$\sigma^{\ast}$ orbitals remain the same as in Fig. 1. Therefore,
the harmonic ellipticity as a probe of the orbital structure for
the nonsymmetric orbitals is also applicable to the multi-cycle
laser pulses.

\section{Conclusion}

We investigate the influence of the orbital symmetry on the
ellipticity of the high harmonics for the nonsymmetric molecular
states. It is found that the ellipticity maps of the high-order
harmonics have distinct shapes for the molecular orbitals with
different symmetry. To gain a deeper insight into the features of
the high-harmonic ellipticity, the structural analysis is
performed for the molecular orbitals. The nodal structure of the
molecular orbital is found to be imprinted on the high-harmonic
ellipticity and the feature of the high-harmonic ellipticity is
essentially determined by the nodal structure of the molecular
orbital. We also investigate the individual contributions of s and
p atomic states to the ellipticity of the high-order harmonics for
the $\sigma$ and $\sigma^{\ast}$ orbital. The results indicate
that the ellipticity of the high-order harmonics can be used as a
probe of the orbital structure for polar molecules

\section*{Acknowledgment}
This work was supported by the National Natural Science Foundation
of China under Grants No. 60925021, 10904045, 11104092 and the
Doctoral fund of Ministry of Education of China under Grant No.
20100142110047.

\end{document}